\documentclass[review]{elsarticle}

\usepackage{amssymb}
\usepackage{graphicx}
\usepackage{lipsum}
\usepackage{float}
\usepackage{color}
\usepackage{lineno}
\usepackage{amsmath}

\journal{International Journal of Heat and Mass Transfer}
\begin{document}
\begin{frontmatter}
 \title{Effects of surface nanostructure and wettability on pool boiling: A molecular dynamics study}
 \author[label1]{Armin Shahmardi\corref{cor1}}
 \cortext[cor1]{Corresponding author: arminsh@mech.kth.se}
\ead{arminsh@mech.kth.se}
\author[label1]{Outi Tammisola}
\author[label2]{Mauro Chinappi}
\author[label1]{Luca Brandt}
 \address[label1]{Linn\'e Flow Centre and SeRC (Swedish e-Science Research Centre), KTH Department of Engineering Mechanics,SE 100 44 Stockholm, Sweden}
 \address[label2]{Dipartimento Ingegneria Industriale, Universit\`a degli studi di Roma Tor Vergata, via del Politecnico 1, 00133, Roma, Italia}
\begin{abstract}
We study the role of surface topology, surface chemistry,  and wall superheat temperature on the onset of boiling, bubble nucleation and growth, and the possible formation of an insulating vapour film by means of  large-scale MD simulations. In the numerical experiments, we control the system pressure by imposing a constant force on a moving piston. 
The simulations reveal that the presence of a nanostructure triggers the bubble formation, determines the nucleation site and facilitates the energy transfer from the hot substrate to the water. 
The surface chemistry, on the other hand,  governs the shape of the formed bubble. A hydrophilic surface chemistry accelerates the bubble nucleation, however, decelerates the bubble expansion, thus postponing the formation of the film of vapour. Therefore,  a hydrophilic surface provides better energy transfer from the hot wall to the water.  By analysing the system energy, we show that irrespective of wall topology and chemistry, there is a wall temperature for which the amount of transferred energy is maximum.
\end{abstract}
\end{frontmatter}


\section{Introduction}
Bubble nucleation and pool boiling heat transfer proved to be of great importance in many industries demanding fast and efficient heat transfer from a hot surface (solar energy, thermal power, microfluidic devices, microelectronics and nanoelectronics, to name a few \cite{intro1,intro2}). Therefore, during the last decades, different experiments 
and numerical simulations based on continuum formulations were performed to understand the physics behind bubble nucleation, evaporation/condensation and boiling \cite{intro3,intro4,intro5,magaletti2015shock,magaletti2016shock,intro02}. 

Li et. al \ \cite{intro0}  performed experiments to study the effects of nanostructure surface treatments on boiling performance at low superheat temperatures. They observed enhanced boiling performance due to the formation of nanobubbles 
induced
by stable nucleation sites at microscale cavities. Zupan\v{c}i\v{c} et. al.\ \cite{intro01} studied nucleate boiling on stainless steel foils by visualising nucleating bubbles and temperature fields using high speed video recording. They showed that bubble nucleation on a flat surface requires higher activation temperature than on nanostructured surface. They also reported that the bubbles on a flat surface are larger than those formed on a nanostructured surface. Shen et. al.\ \cite{intro02}
  employed a diffuse interface model to study bubble growth on a biphilic surface and noticed that, at low gravity, the contact line propagation closely follows the bubble growth everywhere but at the borders between hydrophilic and hydrophobic sections.  However, at high gravity, the bubble expansion becomes weaker and
   the contact line becomes almost stationary at the borders of hydrophilic and hydrophobic sections.

More recently, molecular dynamics (MD) simulations have emerged as a powerful tool to gain detailed information about the physics at the nanoscale also for the case of pool boiling heat transfer. 
Among the available MD studies, Mao and Zhang \cite{intro6}  studied  rapid boiling of a film of water on a hot  surface. These authors observed a rapid phase transition of water molecules close to the surface due to the overheating and  reported the formation of a  constant density non-vaporisation molecular layer attached to the surface of the plate.  The effect of the thickness of the liquid film on the phase transition mechanism (evaporation or  explosive boiling) was examined by Rabbi et al.\ \cite{intro8} by  means of MD simulations of liquid argon over a hot wall.  According to these results, phase change occurs by evaporation for the two thinner films, whereas  the two thicker films undergo explosive boiling. Gupta et al.\ \cite{intro11} studied the onset of  bubble nucleation on a partially heated surface  by  MD simulations and experiments. The effect of the width and temperature of the heated part of the surface on the bubble growth were explored and a critical radius of nucleation reported. These authors also proposed  an analytical model to predict the critical width of the heated part of the surface which would provide  bubble nucleation.

In order to increase the efficiency of the heat transfer process, numerous studies were conducted to design the optimum surface properties by changing the surface chemistry. Hens et al. \cite{intro9} investigated bubble nucleation and film boiling for different superheat temperatures on surfaces with different chemistry  (wettability conditions). These
authors reported that surfaces with hydrophilic chemistry facilitate bubble nucleation or film formation. Zhou et al.\ \cite{intro10} studied bubble nucleation over a biphilic surface and  observed that the nucleation site moves from the hydrophobic to the hydrophilic part as the superheat temperature increases.  Rapid boiling on surfaces with uniform and patterned wettability  was studied by Wu et al.\  \cite{intro12}. Their results show that by increasing the hydrophilic degree of the surface, the water temperature increases and the evaporation rate decreases.

As an alternative strategy, changing the topology of the surface can also affect the heat transfer process.  Fu et al.\  \cite{intro13}   employed cone-shape nanostructured surfaces to investigate the effects of the size of these patterns  on the  rapid boiling of a thin water film by means of  MD simulations. These auhtors showed  that the nanostructures not only increase the heat transfer from the solid substrate but also affect the temperature history and density distribution. Mukherjee et al.  \cite{intro14}  also 
performed MD simulations to study bubble nucleation of liquid water over a silicon solid substrate, focusing on the effect of the nanostructure height, width and type on the bubble growth rate. Zhang et al.\  \cite{intro15} compared the incipient nucleation time and the temperature corresponding to the onset  of boiling of liquid argon over three different nanostructured surfaces, namely flat, concave, and convex. Their MD simulations results  indicate that nanostructured surfaces intensifiy the bubble nucleation. Moreover the same authors reported that  bubble nucleation occurs sooner on a  concave nanostructured surface. Zhang et al.\  \cite{intro7} 
show by means of MD simulations
that the presence of  nanochannels improves the heat transfer from the solid substrate to the liquid argon and intensifies explosive boiling.

Finally,  several studies also considered the combined effect of surface chemistry and topology aiming to better control the onset of boiling, bubble nucleation site,  boiling heat flux, and the formation of the insulating film of vapour. In particular, phase change of an argon liquid over a nanostructured biphilic substrate was studied by Chen et al.~\cite{intro15} whereas Diaz and Guo \cite{intro17} conducted boiling simulation of liquid argon placed on a horizontal  substrate attached to vertical pillars. Measuring the critical heat flux when varying the pillar arrangement (particularly distance) and surface wettability it was concluded that the critical heat flux increases when increasing the distance between the pillars or increasing the degree of hydrophilic chemistry of the surface (i.e.\ decreasing the contact angle).  

It is well known that the bubble nucleation, the temperature of boiling onset,  and boiling heat transfer are affected by the pressure of the system as well as by  the superheat temperature. While the dependence on different superheat temperatures and surface properties  has been studied extensively also at the nanoscale,  most of the previous studies do not consider a  mechanism to control the pressure. 
In particular, all the cited works employ MD simulations with fixed volume. 
This means that the change of the fluid temperature induced by the heat transfer
results in a change of the pressure. 
On the other hand, Marchio et al.\ \cite{Marchio2018} showed that 
standard approaches to perform MD simulations at constant pressure 
(the so-called NPT runs) provide results which 
depend on the size of the system when applied to vapour nucleation.  
It is therefore important to employ a new strategy to control the pressure in MD simulations of boiling systems.

 The goal of this paper is to study bubble nucleation, the formation of the vapour  film, and the energy transfer in a pool boiling simulation under controlled pressure when varying the 
 superheat temperature, surface topology and surface chemistry.
 To properly control the pressure of the system,  we choose to mechanically control the pressure by placing a piston above the slab of water as introduced in~\cite{Marchio2018}.   Furthermore, due to the periodicity of the system, the forming bubble usually grows and merges with its periodic image 
 and generates a vapour film at the wall. In order to minimize the effect of the system size on the results, our system is chosen as large as possible given the current computational constraints and the need to explore different superheat and wetting conditions; it consists of more than one million atoms with a substrate area of about $641$ nm$^2$.  
We will present MD simulations of boiling  water  for different superheat temperatures and over four different solid substrates:
two different kinds of surface chemistry (corresponding to a hydrophilic  and a hydrophobic wall)  and two different topologies of the solid substrate (a flat wall and a wall with a nanocavity).

\section{System setup and simulation method}

\begin{figure} 
\begin{center}  
\includegraphics[width=0.6\textwidth]{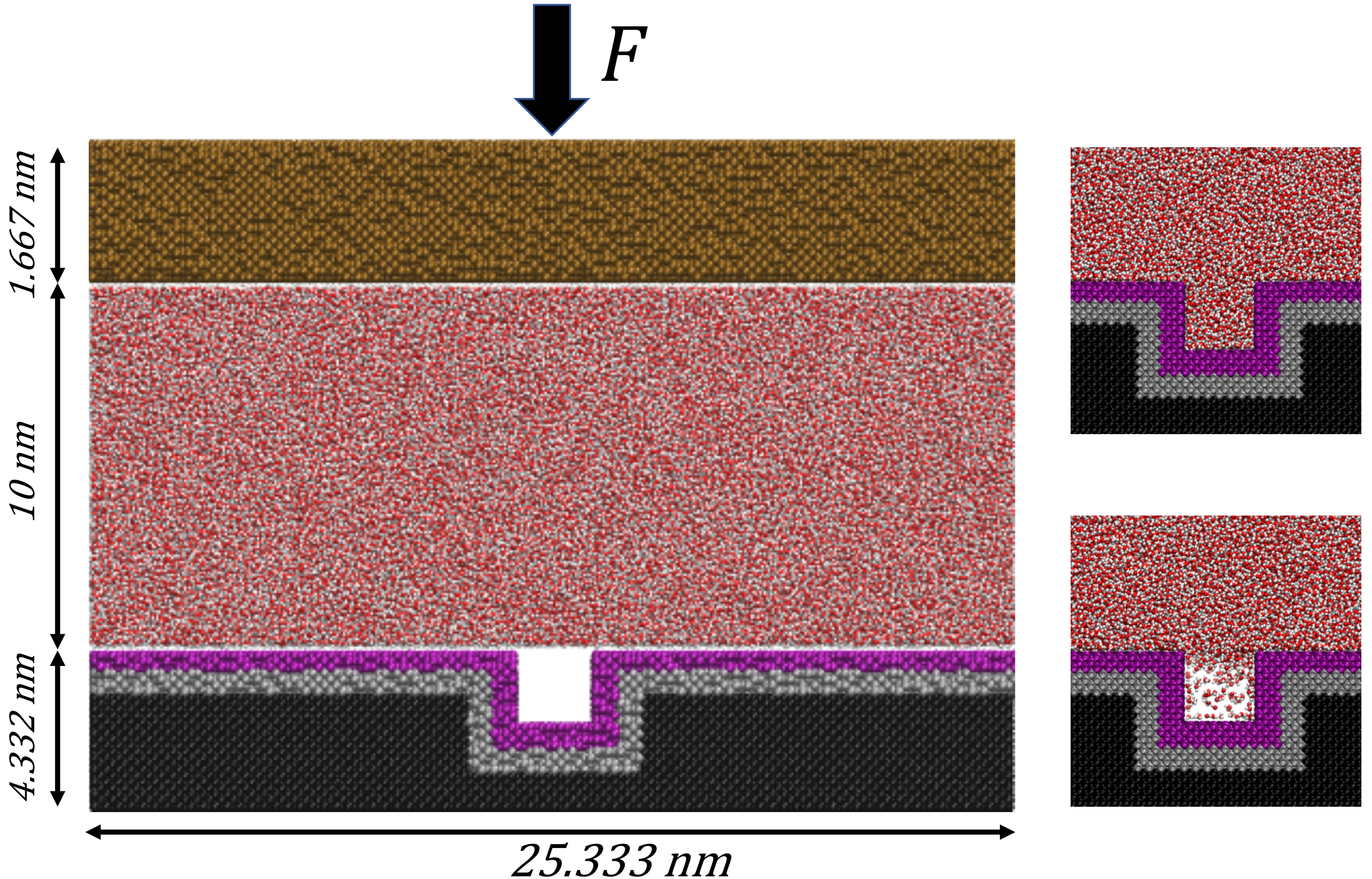}    
\put(-190.0,115){\large a}
\put(-45.0,115){\large b}
\put(-45.0,58){\large c}
\caption{Simulation set-up. a) The simulation system is composed of three components: a solid substrate, a water slab, and a solid piston. The solid substrate consists of three layers, namely the frozen layer (black), the thermostat layer (grey), and the free layer (purple). The components and the layering are the same for the flat substrate for which the cavity is also filled with solid atoms.  For nanostructured wall, the stationary state reached at the end of the equilibration phase can be either Wenzel or Cassie-Baxter state. For hydrophilic cavity, we get a Wenzel state (cavity completely filled by the liquid, panel b) whereas for hydrophobic cavity, a Cassie-Baxter state (liquid does not fill the cavity) is observed, panel c. 
}
\label{fig:System}
 \end{center}
 \end{figure}

The simulated systems consist of three main parts: a solid substrate, a water slab, and a solid piston, see figure~\ref{fig:System}.
We consider both flat and nanostructured substrates. 
In both cases, the substrate consists of atoms arranged in an FCC lattice with a lattice parameter equal to $0.33$~nm. 
Figure~\ref{fig:System} displays the computational setup for the case of the nanostructured substrate,
where a single cavity is present. The solid substrate has dimensions 
$25.33~\mathrm{nm} \times 25.33~\mathrm{nm} \times 4.33~\mathrm{nm}$
and is composed of three layers: 
a bottom fixed layer (black), a thermostated layer (grey) and a free layer (purple).  In the case of the nanostructred wall, the width and height of the cavity is $0.08$ times those of the solid wall.
The interactions among the solid atoms are described by the Lennard-Jones (LJ) potential:
\begin{equation} \label{Lennar-Jones}
\begin{aligned}
\begin{gathered}
V_{ss}
= 
4 \epsilon_{ss} 
\left[ \left(\frac{\sigma_{ss}}{r}\right)^{12}-\left(\frac{\sigma_{ss}}{r}\right)^6 \right],
\end{gathered}
\end{aligned}
\end{equation}
where $\sigma_{ss} = 0.216$~nm is the distance at which the intermolecular potential between the two atoms is zero and $\epsilon_{ss} = 40$~KJ/mol is the depth of the potential well. The LJ parameters are selected  such that the maximum temperature of the simulations ($700$~K) is below the melting temperature of the solid substrate.
All the LJ interactions are cut beyond a cut-off distance $r_c=0.9$~nm. The positions of the substrate (black in the figure) atoms are frozen. During the non-equilibrium simulations, atoms in the thermostated (purple) region are restrained to their initial lattice positions through a harmonic potential:
\begin{equation} \label{PositionRestrain}
\begin{aligned}
\begin{gathered}
V_{pr}(\boldsymbol{r_i})= \frac{1}{2} k_{pr} {(\boldsymbol{r_i}-\boldsymbol{R_i})}^2,
\end{gathered}
\end{aligned}
\end{equation}
where $\boldsymbol{r_i} $ is the position of atom $i$ at  time $t$, $\boldsymbol{R_i} $ is the initial position of atom $i$, and $k_{pr} = 10^3\,\mathrm{KJ/(mol\,nm^2)}$ is the spring constant. 
Moreover, atoms in the thermostated region are connected to a velocity rescale thermostat~\cite{Holian1995}.

To properly control the pressure, a piston (brown atoms in the figure \ref{fig:System})  is placed on top of the water slab. This is free to move up and down thus providing a mechanical control of the pressure as introduced in~\cite{Marchio2018}. The piston is also modelled as an FCC solid with same lattice parameter as the substrate. The height of the piston is  $3.66~\mathrm{nm}$. A constant downward acceleration is imposed on all the atoms of the piston, with magnitude  computed  to provide the  force on the piston which corresponds to the prescribed pressure.
In particular, being $F$ the total downward force acting on the piston,  $f$ the downward force on each atom, $A$ the area of the piston and $n_p$ the number of atoms of the piston, we have:
\begin{equation} \label{acceleration}
P 
= 
\frac{F}{A} 
= 
\frac
{f n_p}
{A} \, .
\end{equation}
Finally, the SPC/E model \cite{Berendsen1987} is used for water.
The height of the water slab is $\simeq 10$~nm, with
periodicity assumed in all the three directions. 
The simulation box is high enough ($200$~nm) to ensure that during the boiling 
non-equilibrium simulations, the motion of the piston is not affected by the periodic image of the solid substrate. For the single cavity reported in figure~\ref{fig:System}, the system contains around $1.2$ million atoms including (approximately) $190$K atoms in  the frozen layer, $54$K  atoms connected to the thermostat, $62$K atoms in the free layer of the solid wall, $630$K atoms in water slab, and $250$K atoms forming the piston.

\subsection{Wettability of the substrate}
\begin{figure} 
\begin{center}  
\includegraphics[width=0.8\textwidth]{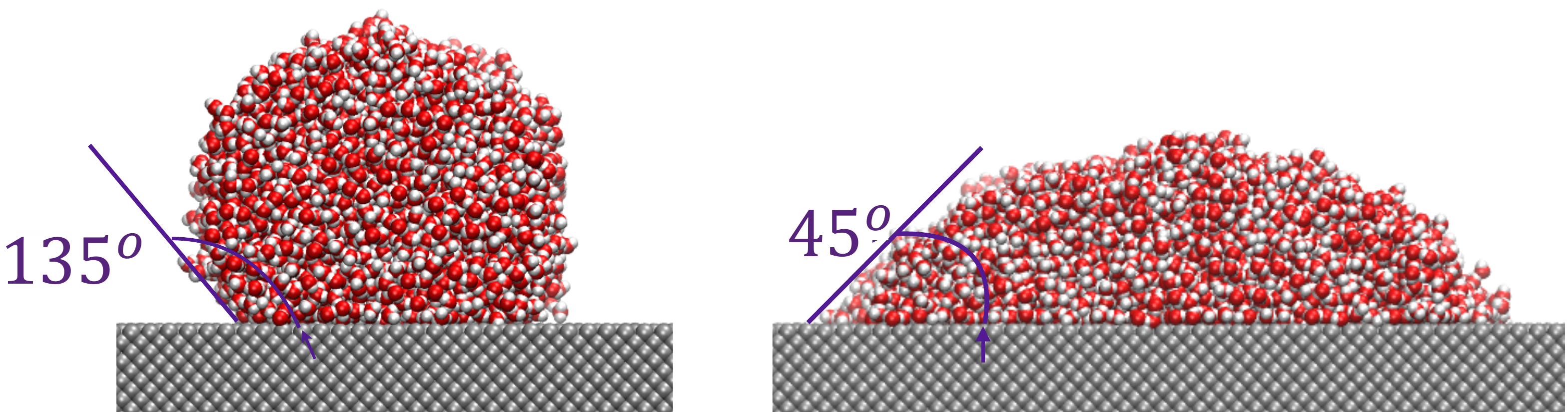}      
\caption{Wettability simulations. Droplet equilibrium over a hydrophobic and hydrophilic wall. The results show that the contact angles for the hyrdophobic and hydrophilic walls are about $135^{\circ}$ and $45^{\circ}$.
}
\label{fig:Wettability}
 \end{center}
 \end{figure}

The wettability of the substrate is controlled by the water-substrate interaction 
potential. The oxygen atoms of water molecules interact with solid atoms 
via a Lennard-Jones potential
\begin{equation} \label{Lennar-JonesWater}
\begin{aligned}
\begin{gathered}
V_{so}= 4\epsilon_{so} \left[ \left(\frac{\sigma_{so}}{r}\right)^{12}-\left(\frac{\sigma_{so}}{r}\right)^6 \right],
\end{gathered}
\end{aligned}
\end{equation}
\sloppy
where $\sigma_{so}$ is equal to the arithmetic average of $\sigma_{ss}=0.216$~nm and $\sigma_{oo} = 0.317$~nm, according to the Lorentz-Berthelot rule \cite{Nagayama2010}.
The value of $\epsilon_{so}$ controls the wettability. Several wetting simulations were performed on a cylindrical droplet\ \cite{cylindrical} for different values of  $\epsilon_{so}$ to obtain the values corresponding to a hydrophobic and a hydrophilic surface. In particular, we have selected  $\epsilon_{so}=0.55$ ~KJ/mol for an hydrophobic substrate
(contact angle $\theta \approx 135^\circ$) and $\epsilon_{so}=1.4$~KJ/mol 
for the hydrophilic substrate (contact angle $\theta \approx 45^\circ$). Figure \ref{fig:Wettability} presents the equilibrium configuration from the wetting simulations over the hydrophilic and hydrophobic flat walls. Note that the contact angles are measured graphically and the reported angles are approximated values which are accurate enough for our purpose (providing a hydrophilic and a hydrophobic surface). 

\subsection{Simulation protocol}

All the simulations are performed using the open-source software GROMACS~\cite{GROMACS}.
The simulation protocol is as follows. First, each of the three components of the system (solid substrate, water slab, and piston) are separately equilibrated for $1$~ns. The solid substrate is equilibrated at $T=300$~K using an NVT simulation. To this aim, all the three layers are connected to a velocity rescaling thermostat for $1$~ns. The piston is also equilibrated at $T=300$~K using the same thermostat.
Finally, the water slab is  equilibrated using an NPT semi-isotropic simulation where a Parrinello-Rahman barostat \cite{ParrinelloRahman1981} is used together with a velocity rescaling  thermostat to equilibrate the water box at $T=300$~K and $P=1$~bar.

Secondly, the equilibrated components are merged as shown in figure~\ref{fig:System}. 
An additional NVT equilibration is then performed on the integrated system for $1$~ns. During this run, the piston is active, i.e. a constant acceleration (or equivalently a constant force) is applied on each atom forming the piston.
Depending on the magnitude of the mechanically applied pressure and on the chemistry of the surface, the stationary state reached by the system after this equilibration phase can be either a Cassie-Baxter or a Wenzel state~\cite{cassie1944wettability,Wenzel,Giacomello,meloni2016focus}. For the hydrophilic nanostructure cavity we get that the system spontaneously moves to a Wenzel state (the cavity is completely filled by the liquid) whereas the stationary state for a hydrophobic nanostructured wall is a Cassie-Baxter state (the liquid does not fill the cavity).  
Finally, the thermostat is disconnected from all the atoms except those in thermostat group of the solid substrate (see figure~\ref{fig:System})
and the non-equilibrium molecular dynamics simulations are started. 

As introduced above, we shall study two substrates, namely, a flat wall and a nanostructured wall. For each substrate, two different wettabilities are studied, hydrophobic and hydrophilic. Seven thermostat temperatures are considered, $T_W=$ $400$~K,  $450$~K, $500$~K, $550$~K, $600$~K, $650$~K, and $700$~K. For all the simulations we use a time step of $0.1$~fs except for those at temperature larger than $550$~K when the time step is reduced to $0.05$~fs. For the sake of computational cost, the simulations are limited by two criteria. i) Simulations are performed up to maximum 8~ns, or ii) they are stopped when the vapour film completely covers the wall, because we are only interested in the physics prior to the formation of the vapour film, i.e.\ when the solid substrate is not insulated by the vapour film. 

 \section{Results}
The results of the simulations of boiling on surfaces with different topologies and wettabilities will be analysed in terms of
bubble formation and shape, location of the nucleation site, bubble growth, film formation, and energy transfer from the hot wall to the water slab and the piston.


\subsection{Bubble nucleation and bubble growth}  \label{sec:nucleation}
\begin{figure} [t!]
\begin{center}  
\includegraphics[width=1.\textwidth]{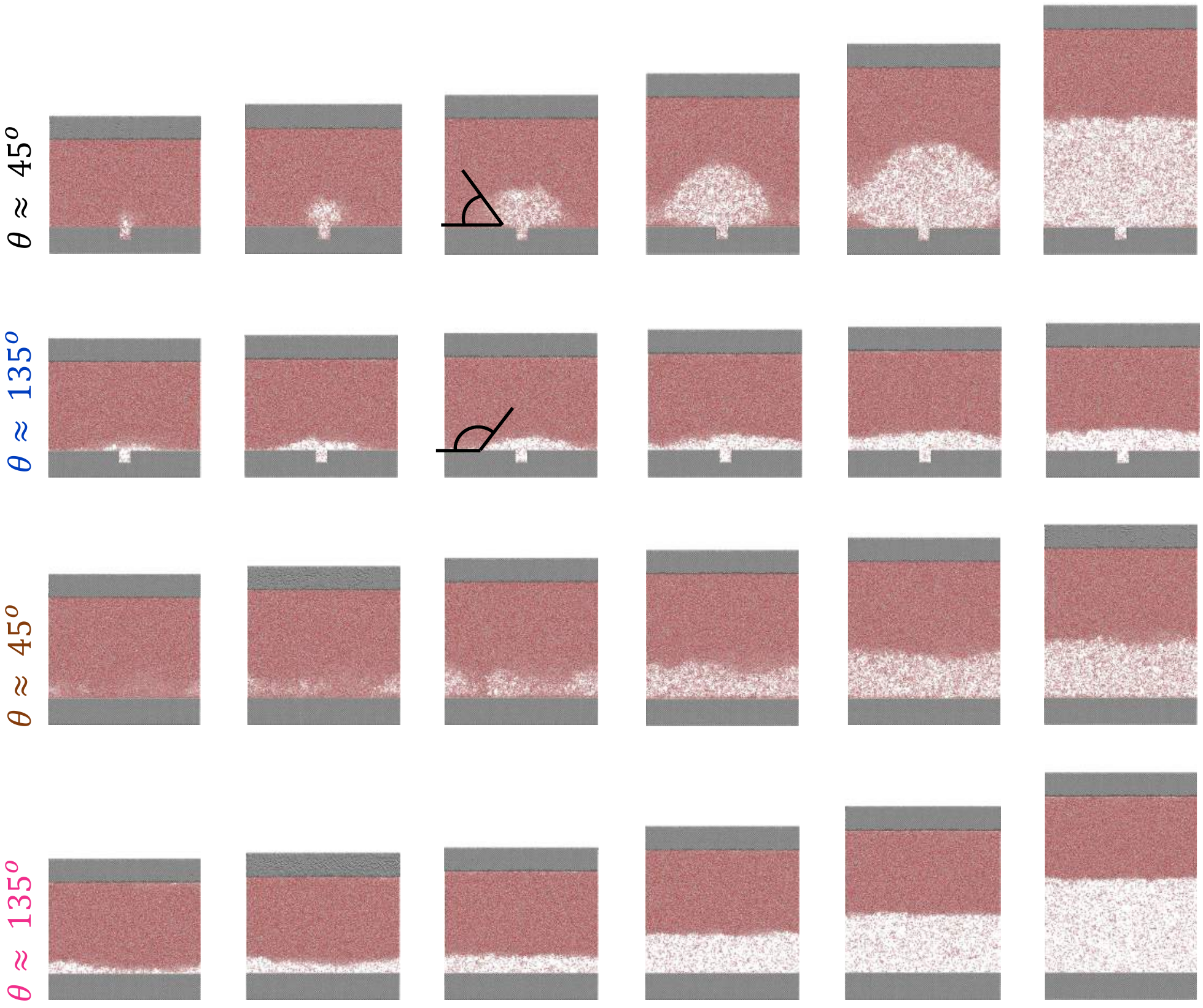}
\put(-330.0,207){\scriptsize a,t$=1.325$~ns}
\put(-270.0,207){\scriptsize  b,t$=1.45$~ns}
\put(-216.0,207){\scriptsize c,t$=1.575$~ns}
\put(-156.0,207){\scriptsize d,t$=1.7$~ns}
\put(-103.0,207){\scriptsize e,t$=1.825$~ns}
\put(-42.0,207) {\scriptsize f,t$=1.95$~ns}
\put(-330.0,142){\scriptsize g,t$=4.85$~ns}
\put(-270.0,142){\scriptsize h,t$=4.9$~ns}
\put(-216.0,142){\scriptsize i,t$=4.95$~ns}
\put(-156.0,142){\scriptsize j,t$=5.0$~ns}
\put(-103.0,142){\scriptsize k,t$=5.05$~ns}
\put(-42.0,142) {\scriptsize l,t$=5.1$~ns}
\put(-330.0,72){\scriptsize m,t$=1.625$~ns}
\put(-270.0,72){\scriptsize n,t$=1.675$~ns}
\put(-216.0,72){\scriptsize o,t$=1.725$~ns}
\put(-156.0,72){\scriptsize p,t$=1.775$~ns}
\put(-103.0,72){\scriptsize q,t$=1.825$~ns}
\put(-46.0,72){\scriptsize r,t$=1.875$~ns}
\put(-330.0,-7){\scriptsize s,t$=4.625$~ns}
\put(-270.0,-7){\scriptsize t,t$=4.7$~ns}
\put(-216.0,-7){\scriptsize u,t$=4.775$~ns}
\put(-156.0,-7){\scriptsize v,t$=4.85$~ns}
\put(-103.0,-7){\scriptsize w,t$=4.925$~ns}
\put(-42.0,-7){\scriptsize x,t$=5.0$~ns}
\caption{Evolution of the nucleated bubble for different wall chemistries and topologies.
The thermostat temperature is set to $T_W=700$~K. The first and the second rows correspond to nano-structured walls while the third and the fourth rows show the evolution of the bubble (or  the film) on a flat wall. The walls in the first and the third rows are hydrophilic whereas the walls in the second and the fourth rows are hydrophobic. The presence of a nanostructure mostly controls the nucleation site, whereas the bubble shape (contact angle) depends on the surface chemistry.}
\label{fig:BG}
 \end{center}
 \end{figure}
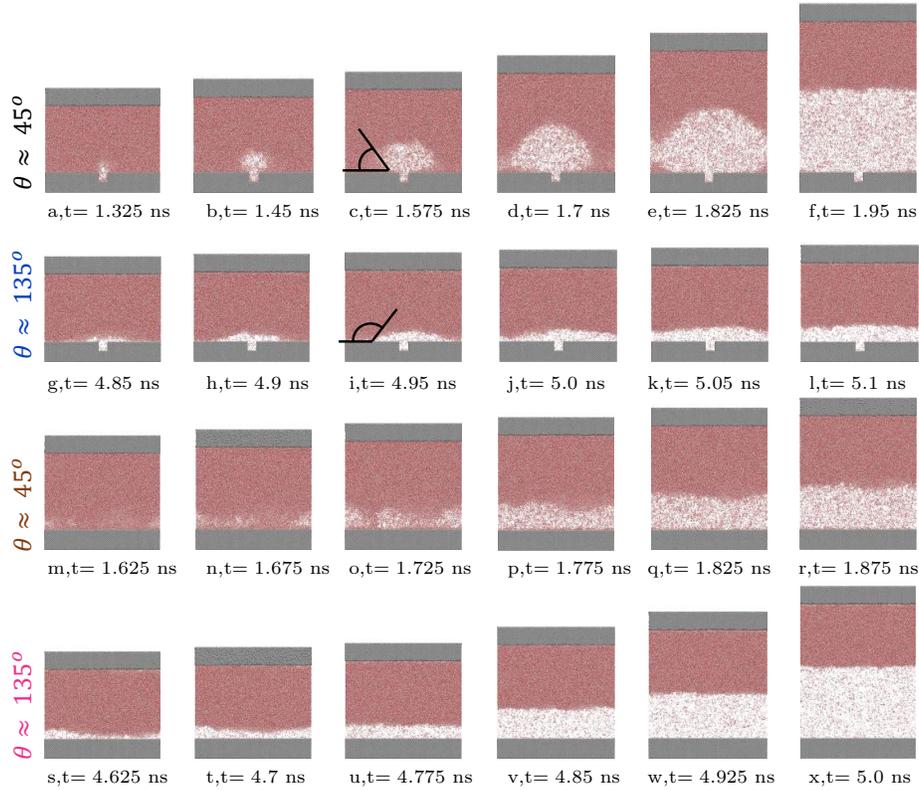

The formation, the shape, and the growth of the bubble or the film of the vapour strongly depend on the topology and the chemical properties of the wall. Figure~\ref{fig:BG} displays the evolution of the nucleated bubble for the case with the thermostat temperature  of  $700$~K (hereafter this will be referred to as wall temperature and denoted by $T_W$) for  the 4 different wall chemistries and  topologies under consideration. 
 
For a hydrophilic structured wall (fig~\ref{fig:BG}a-f) the initial state of the system after the equilibration is the Wenzel state, i.e.\ the water completely occupies the cavity \cite{Wenzel}.  The nucleation occurs in the cavity at 
$t\simeq1.325~ns$ (fig~\ref{fig:BG}a). The bubble grows gradually in the vicinity of the cavity (fig~\ref{fig:BG}b-d) until it merges with its periodic image (fig~\ref{fig:BG}e) and forms a film of water vapour (\ref{fig:BG}f). During the growth, the contact angle is $\sim 45^\circ$, which accommodates the hydrophilic chemistry of the wall (see fig~\ref{fig:BG}c).
For a hydrophobic structured wall (fig~\ref{fig:BG}g-l), the initial state of the system is the Cassie--Baxter state~\cite{cassie1944wettability},  the bubble nucleates in the cavity as for the hydrophilic structured wall (fig~\ref{fig:BG}g), although  later ($t\simeq4.85$~ns) than in the case of the hydrophilic structured wall. 
After the nucleation, the bubble  grows (see fig~\ref{fig:BG}h-i) until it forms the insulating vapour  film  
by merging with its periodic image (fig~\ref{fig:BG}j-l).
 Analogously to the boiling over the hydrophilic structured wall, the shape of the bubble is dictated by the chemistry of the surface. Thus, during the bubble growth, the contact angle is $\sim 135^\circ$ (see fig~\ref{fig:BG}i).
 
\begin{figure}  [t!]
\begin{center}  
\includegraphics[width=0.58\textwidth]{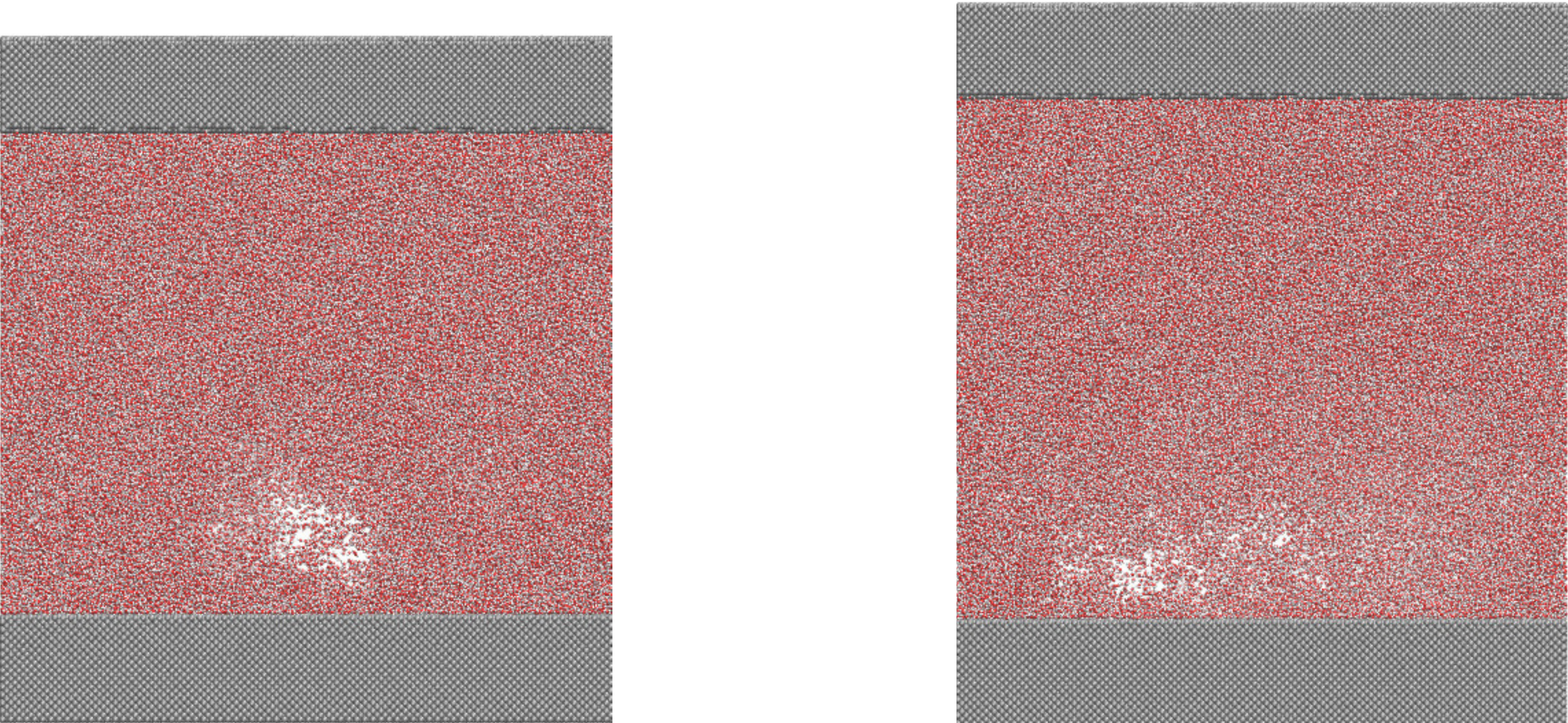} 
\put(-208.0,-7){\scriptsize a) $ T_W=600$~K,t$=3.025$~ns}
\put(-90.0,-7){\scriptsize b) $ T_W=650$~K,t$=2.025$~ns}
\caption{Random nucleation over a heated flat wall. The two panels report the first stage of the nucleation 
for the case of the hydrophilic flat wall, (a) $T_W=600$K, b) $T_W=650$K). The bubbles nucleate  at different locations close to the wall.}
\label{fig:NucleationAll}
 \end{center}
 \end{figure}

A completely different scenario is observed for the cases with the flat walls. In the hydrophilic case (fig~\ref{fig:BG}m-r), the nucleation occurs in a random location in the bulk liquid (fig~\ref{fig:BG}m),  while, for the flat hydrophobic case, there is no unique nucleation site (fig~\ref{fig:BG}s), and the boiling starts with 
 several nucleation sites which quickly merge and form a vapour film  (fig~\ref{fig:BG}t-x). To highlight  the randomness of the nucleation site for a flat hydrophilic wall, the nuclei from the two simulations with wall temperature $T_W=600$~K and $T_W=650$~K, are depicted in  figure~\ref{fig:NucleationAll}.

To summarise, the presence of a nanocavity at the wall triggers the formation of a bubble and determines a unique nucleation site, as also shown  in 
previous MD studies on bubble nucleation over structured surfaces (see for instance Ref.\ \cite{intro15}). On the other hand,  the surface chemistry governs the shape and the growth rate of the forming bubble.  The hydrophilic surface accelerates  the bubble nucleation  whereas  the hydrophobic chemistry postpones  it
as also reported from the MD simulations of liquid argon by Hens et al.~\cite{intro9}.

 \subsection{Energy analysis}
In the following, we will consider the system consisting of the 
water slab and the piston and examine the amount and the rate of energy transfer from the wall to the system, denoted as $\Delta E$ and $\dot{E}$ respectively.
The total energy of the system is computed as the sum of  five different contributions:
\begin{equation} 
\begin{aligned}
\begin{gathered}
E = U_w + K_w + U_p + K_p + U_s
\end{gathered}
\end{aligned}
\label{eq:Etot}
\end{equation}
where 
$U_{w}$ (water potential energy) is the summation of the Lennard-Jones and the Coulomb interaction energy among the water molecules, 
$K_{w}$ is the kinetic energy of the water atoms, 
$U_{p}$ is the Lennard-Jones interaction energy among the piston atoms, 
$K_{p}$ represents the kinetic energy of the piston atoms, and 
finally, $U_{s}$ is the sum of all the surface energies, i.e.\
the Lennard-Jones interaction of the water atoms with the wall  
and the piston atoms.

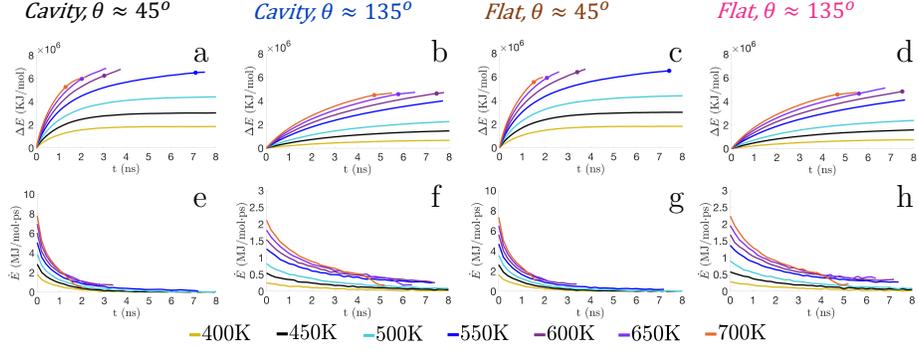
\begin{figure} [H] 
\begin{center}  
\includegraphics[width=1.\textwidth]{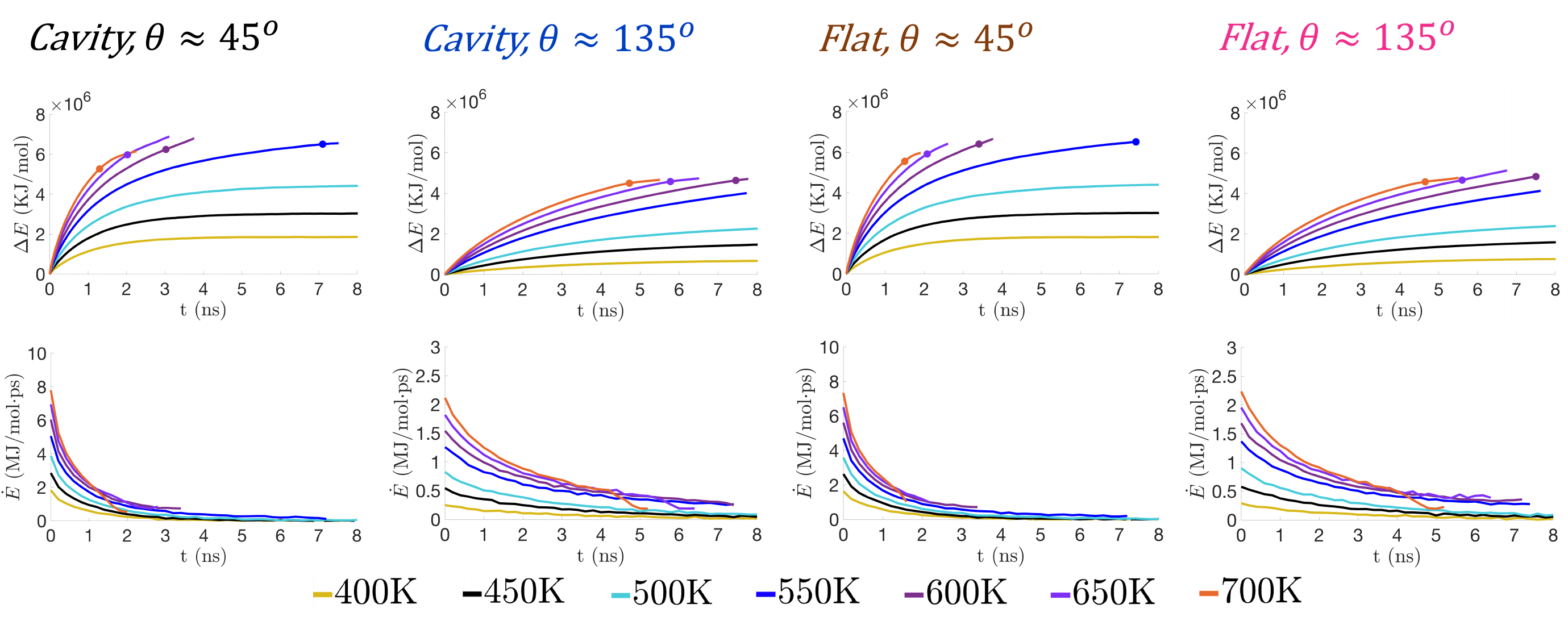} 
\put(-275,110){\large  a}
\put(-185,110){\large  b}
\put(-95,110){\large  c}
\put(-9.0,110){\large  d}
\put(-275,55){\large  e}
\put(-185,55){\large  f}
\put(-95,55){\large  g}
\put(-9.0,55){\large  h}
\caption{Time evolution of the system energy and of its rate of change. Panels a-d depict  $\Delta E= E(t) - E_0$ with $E_0$ the initial energy while panel e-g the  rate of change $ \dot{E} = d E / d t $ for the four combination of 
wall wettability and topologies under consideration. 
The onset of boiling, when observed, is indicated with a filled circle in the energy plot, see e.g. the red circle in panel a), which corresponds to the visualisation in panel a) of figure \ref{fig:BG}. For some cases, boiling does not start during the $8~ns$ of the simulation.
\label{fig:EnRate}}
\end{center}
\end{figure}

The results of the energy analysis are reported in figure~\ref{fig:EnRate}, where we report  the time evolution of the  system total energy,  $\Delta E= E(t) - E_0$ (where  $E_0$ stands for the initial energy of the system), and its rate of change, $ \dot{E} = d E / d t $, for the different wall contact angles and topologies under consideration. 
To reduce the noise, $\dot{E}$ is obtained as an average over time intervals of $200$~ps. 
Some general conclusions can be drawn from the data in  figure~\ref{fig:EnRate}. 
First, both the energy of the system, $E$, and its rate of change, $\dot{E}$, are mostly affected by the change in the chemistry of the surface rather than by the topology of the wall. Because, the interactions between the solid-wall and water molecules are stronger for a hydrophilic wall, the increase of the system energy is faster and more intense over the hydrophilic wall and, consequently, the bubble (or the vapour film) formation  occurs  in a shorter time. 
This observation is consistent with the results in section~\ref{sec:nucleation}  where it was shown that the bubble (or the film) inception time is mainly  affected by the surface chemistry; as example, in the case of the nano-structured topology, nucleation occurs  after $1.325$~ns over a hydrophilic wall (fig~\ref{fig:BG}a) whereas it requires  $4.85$~ns in the case of a hydrophobic wall (fig~\ref{fig:BG}g).

Secondly, $\Delta E$ and $\dot{E}$ increase with the wall temperature before the onset of boiling for all the cases (the time corresponding to the onset of boiling is marked with  symbols in the plots in the first row of fig~~\ref{fig:EnRate}). Note, however, that boiling does not start during the $8~ns$ of simulations  for some of the cases. The reason for the absence of any nucleation (or alternatively vapour film) in these cases will be discussed in  section~\ref{sec:TemperatureAnalysis}.

Finally,  after the onset of boiling, the hot wall is partially insulated by the vapour phase. Therefore, the total system energy reaches a plateau  once the wall is fully covered by the film of vapour and the energy rate of change decreases dramatically.

\begin{figure} 
\begin{center}  
\includegraphics[width=0.8\textwidth]{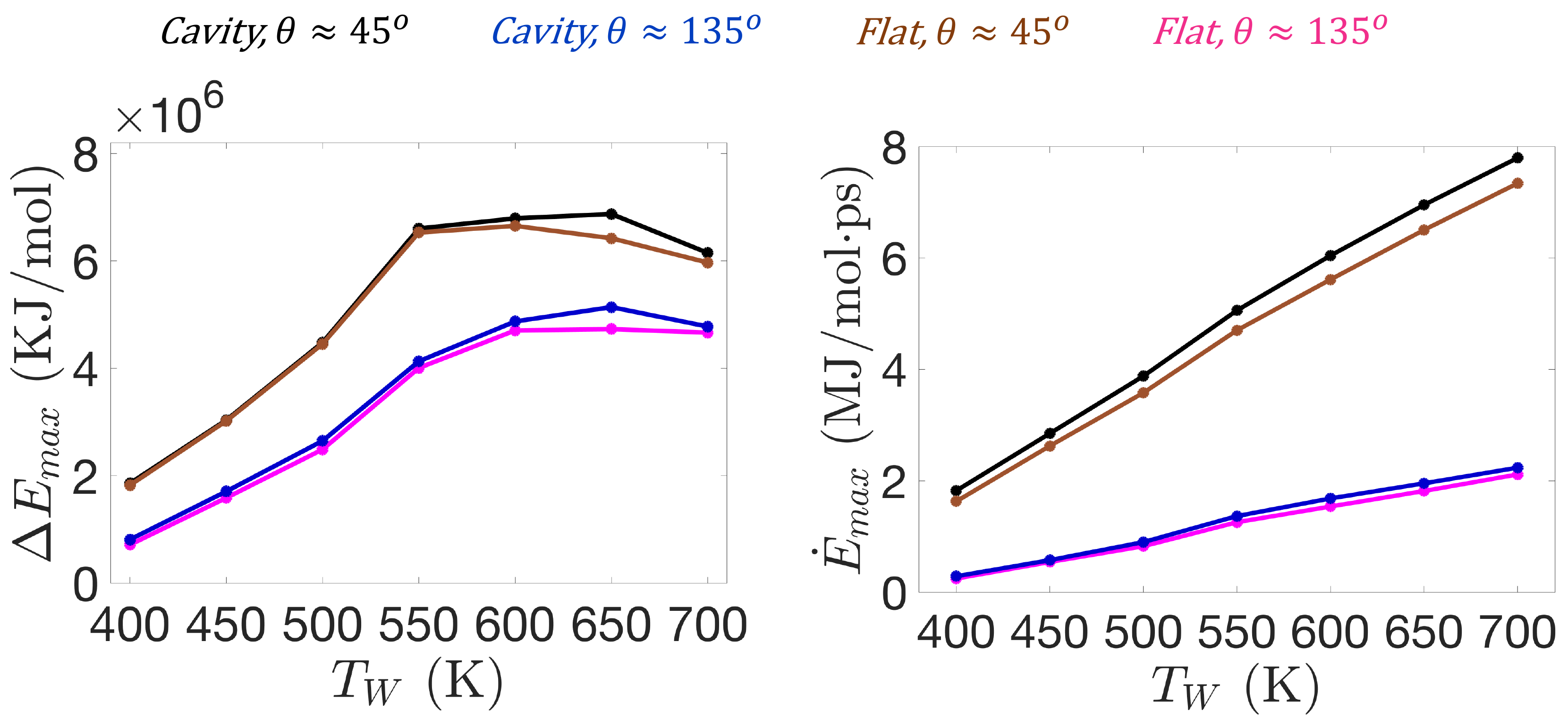} 
\put(-250,85){\large  a}
\put(-105,85){\large  b}
\caption{
\label{fig:BoilCu}
Effect of wall temperature on the energy transfer. a) Maximum amount of energy extracted by the water slab from the hot wall  and b) maximum rate of change of the energy of the system. Although the maximum rate of energy transfer ($\dot{E}_{max}$) increases with the wall temperature, the maximum amount of transferred energy, $\Delta E_{max}$, first increases with wall temperature  and then reduces. }
\end{center}
\end{figure} 

Next, we report in figure~\ref{fig:BoilCu} the effect of different wall temperatures, $T_W$, on the maximum amount of energy extracted from the hot wall by the water slab, $\Delta E_{max}$, and the maximum rate of energy transfer between the hot wall and the system, $\dot{E}_{max}$ (fig~\ref{fig:BoilCu}a and  fig~\ref{fig:BoilCu}b).
As first observation, $\dot{E}$ decreases with time, see fig 5e-h.
This is in agreement with standard continuum intuition 
that suggests that $\dot{E}$ is roughly proportional to the temperature 
difference between the solid wall and the temperature of the portion of the 
water slab in contact to the wall.
This difference is maximum at beginning of the process 
when the water slab is at $300$~K and it decreases as the water 
temperature increases. 
Furthermore, the maximum amount of the energy transferred $\Delta E_{max}$ first increases with the wall temperature  and then  reduces. 
This behaviour is attributed to the combined effect of bubble formation and growth and to the increase of the rate of energy transfer at high temperature. For the highest wall temperatures, the vapour bubble forms sooner and grows faster. 
Thus, an insulating vapour layer partially or fully covers the wall relatively quickly,  which results in a reduction of the energy transfer.   On the other hand, the rate of energy transfer increases at these higher temperatures. These two effects explain the presence of a maximum in the curves in fig~\ref{fig:BoilCu}a.

As mentioned earlier, the chemistry of the wall has a dominant role for the amount of energy transfer to the system. 
However, according to fig.~\ref{fig:BoilCu}a,  given the surface chemistry, the wall temperature at which the amount of transferred energy is maximum differs for the different surface topologies.  For the cases with nano-structured surfaces, the temperature corresponding to the maximum transferred energy is higher than that for the flat walls. Considering the discussion above about the origin of the maximum in the energy transfer curve in fig~~\ref{fig:BoilCu}a, the data confirm that the surface topology also affects the bubble growth and the formation of the film (although less than the surface chemistry). For nano-structured surfaces, the bubble forms and grows around the cavity and  its growth rate  is  therefore limited by this geometrical constraint. This is in accordance with the data in the first and the third rows of fig~\ref{fig:BG}, where it is shown that the transition from a nucleated bubble (fig~\ref{fig:BG}a) to the film formation (fig~\ref{fig:BG}f) takes about $0.625
$~ns for the nano-structured hydrophilic wall, while the same process takes about $0.2$~ns over the flat hydrophilic wall,   see fig~\ref{fig:BG}m-q.  Thus, a nanostructure delays the formation of the vapour film and improves the energy transfer which is in agreement with the results of the experiments performed by Das et. al.  \cite{DAS2017}. Their results illustrate that nanostructured surfaces increase the pool boiling heat transfer by increasing the effective heating surface.

\begin{figure}  
\begin{center}  
 
\includegraphics[width=01.0\textwidth]{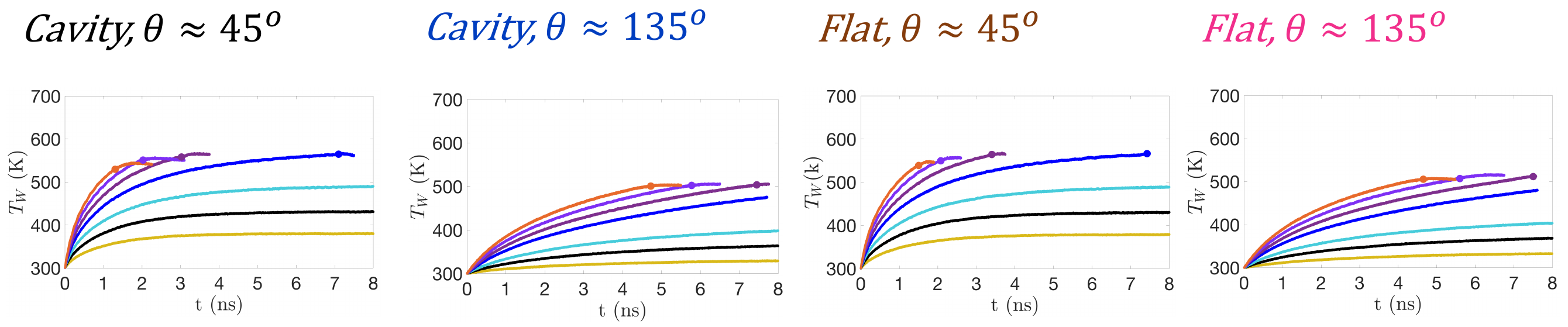} 
\put(-270,44){\large  a}
\put(-185,40){\large  b}
\put(-95,44){\large  c}
\put(-12,40){\large  d}
\caption{Evolution of the average water temperature.  Due to the formation of the insulating vapour film, the averaged water temperature never reaches the thermostat temperature. 
\label{fig:Temperatures}} 
\end{center}
\end{figure}
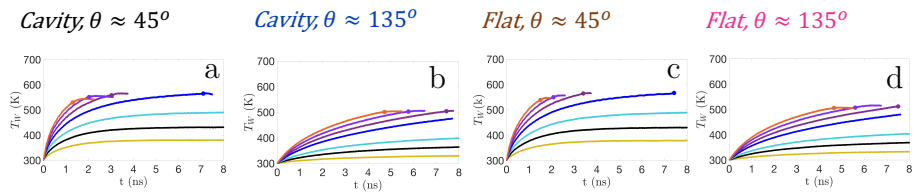 

\subsection{Temperature fields} \label{sec:TemperatureAnalysis}

 The evolution of the averaged temperature of the water is displayed in figure~\ref{fig:Temperatures} 
for the different wall topologies, contact angles and different thermostat temperatures examined. 

 First we note that for all cases, irrespective of the wall chemistry and topology, the averaged water temperature does not reach that of the wall during the $8$~ns of simulation. Let us consider the hydrophilic structured wall (fig~\ref{fig:Temperatures}a) for which the bubble nucleation, bubble growth, and film formation  occur during the simulation time if the wall temperature is greater than $500$~K. 
 For these wall temperatures above $500$~K, the vapour phase insulates the hot wall partially (or fully) before the averaged water temperature reaches that of the wall. Therefore, the  averaged water temperature is always below the wall temperature. 
 
\begin{figure} [H]
 \begin{center}  
 \includegraphics[width=1.0\textwidth]{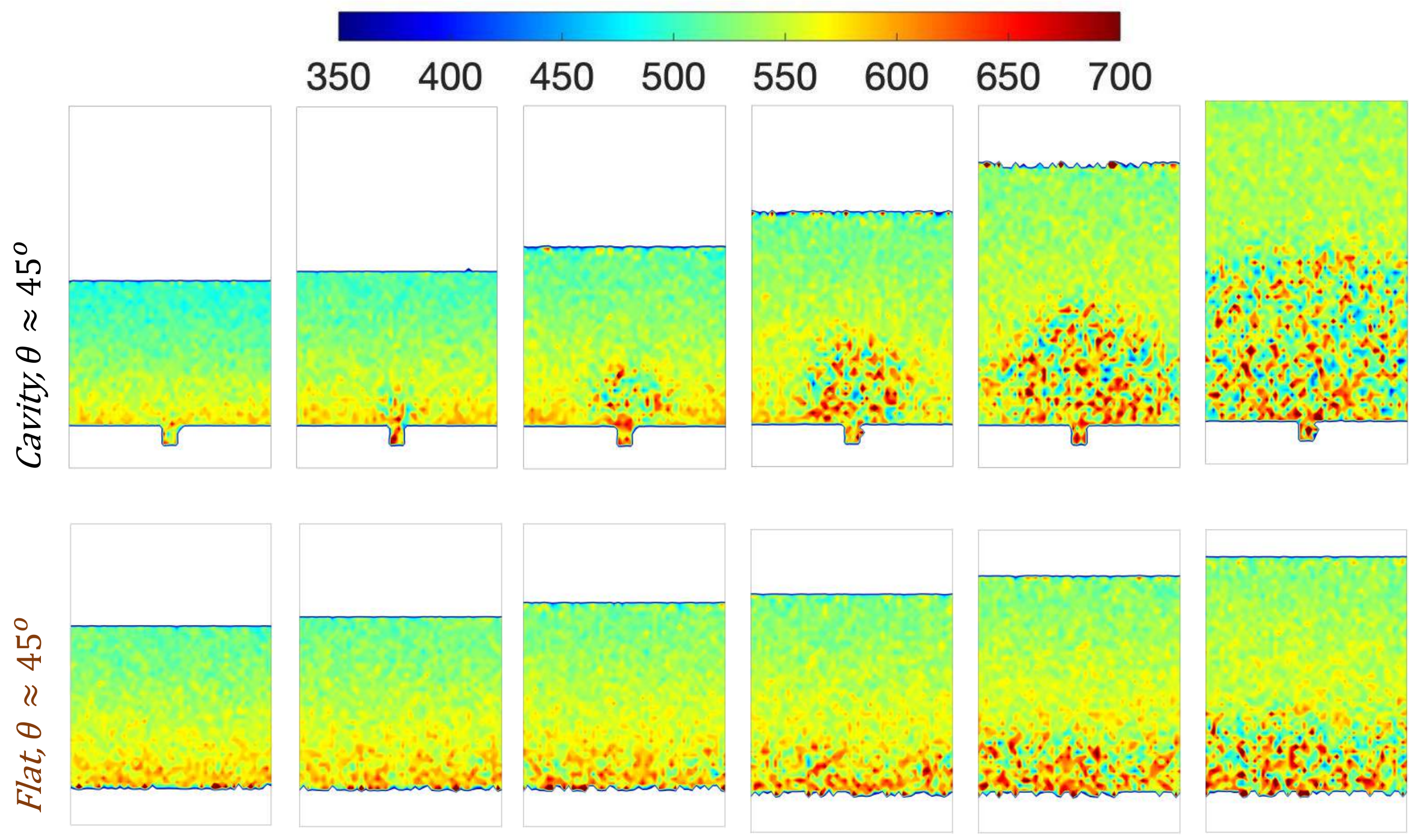} 
   \put(-305,80){\large  a}
   \put(-250,80){\large  b}
   \put(-195,80){\large  c}
   \put(-140,80){\large  d}
   \put(-85,80){\large  e}
   \put(-30,80){\large  f}
   \put(-305,-4){\large  m}
   \put(-250,-4){\large  n}
   \put(-195,-4){\large  o}
   \put(-140,-4){\large  p}
   \put(-85,-4){\large  q}
   \put(-30,-4){\large  r}
 \caption{Time evolution of the water temperature field for hydrophilic cases. $T_W=700$~K.  Although the averaged wall temperature is below that of thermostat, the water temperature reaches the thermostat temperature locally.}
 \label{fig:TempCont}

 \end{center}
\end{figure}
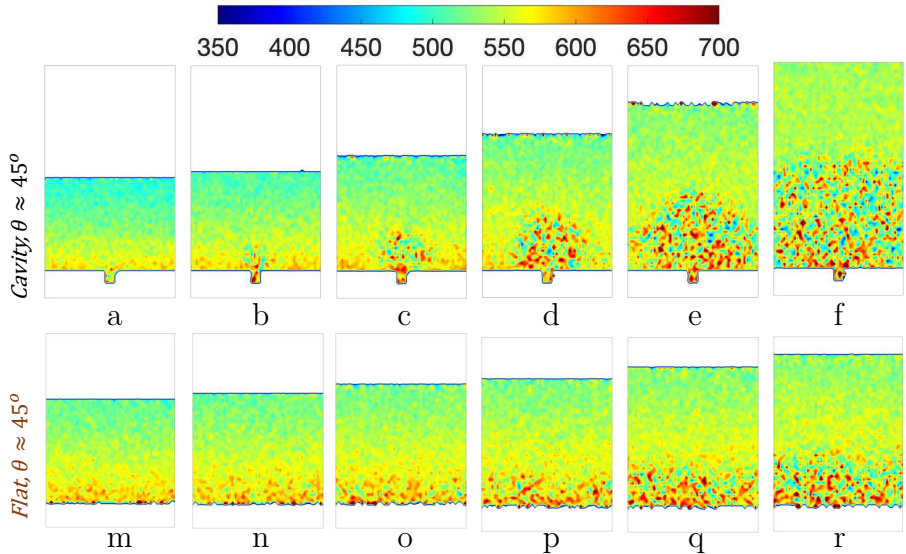

 Next, as discussed earlier,  higher wall temperatures accelerate the onset of boiling and the formation of the insulating layer. As a consequence, in the case where the wall temperature is $700$~K, the difference between the equilibrium water temperature (around $550$~K) and the wall temperature is higher than that at the lower temperatures.  Note also that for the cases  with thermostat temperatures equal and below $500$~K, the average water temperature is still slightly increasing with time, which indicates that longer simulation times would be necessary to reach the equilibrium temperature and observing bubble nulceation.
Although the average water temperature is always below the wall temperature, locally the water temperature does  reach that of the wall. This is documented in 
figure~\ref{fig:TempCont} by iso-contours of local temperature of the water molecules for the boiling over  hydrophilic  walls and thermostat temperature  $700$~K, where the time frames are selected at the same instances as those in figure~\ref{fig:BG}. Indeed, as suggested by the visualisations in  figure~\ref{fig:BG}, the local temperature is higher inside the cavity at the onset of the boiling, where it is close to the wall temperature (see fig~\ref{fig:TempCont} a) and where
 nucleation is therefore seen. 
As time evolves, the region of high local temperature expands on the upper surface  (cf.\ fig~\ref{fig:TempCont}b-f), leading to bubble growth and the formation of the film, as also discussed earlier. Similarly, at the onset of boiling over the flat hydrophilic wall, the local temperature  is highest at the nucleation site, close to the wall temperature \ (see over the left part of the wall in fig~\ref{fig:TempCont} m).

Finally, let us consider  fig~\ref{fig:Temperatures}a and fig~\ref{fig:Temperatures}c  (or  fig~\ref{fig:Temperatures}b and fig~\ref{fig:Temperatures}d). The results reveal that for a given surface chemistry and wall temperature, the averaged temperature of the system corresponding to the onset of boiling (filled circles on the plots) is greater if the wall is not nano-structured.  This observation is in agreement with the experimental study reported by Zupan\v{c}i\v{c} et al.\ \cite{intro01}. Their results show that the bubble nucleation on a flat surface requires higher activation temperature than on a nanostructured surface.
\section{Conclusion}
The onset of boiling, bubble nucleation and growth, and the possible formation of an insulating vapour film are investigated by means of  large-scale MD simulations and analysing the system energy evolution. In particular, we consider a hydrophilic and hydrophobic wall, with corresponding contact angles of 45 and 135 degrees, and two wall topologies, a flat wall and a wall with a periodic array of nano-cavities, and vary the temperature of the solid substrate from  $400$~K to $700$~K.
Specific and novel to this set of simulations is the control of the system pressure by means of a piston on the top boundary. A downward constant force is imposed on the piston atoms providing a mechanical control of the averaged pressure of the system at the desired value  ($P=1$~bar).

The results of the simulations with different wall topologies reveal that the presence of a nanostructure triggers the bubble formation and determines the nucleation site. The formed bubble expands around the nanostructure, thus the bubble growth is geometrically controlled by the cavity. This geometrical control slows down the bubble expansion and the formation of the vapour film which insulates the wall. Therefore, the presence of a nanostructure facilitates the energy transfer from the hot substrate to the water by controlling the nucleation site, detaining the bubble growth, and postponing the formation of the vapour film.

A concerns the wall chemistry, the results indicate that the value of the contact angle determined the 
the shape of the formed bubble. A hydrophilic surface accelerates the bubble nucleation, however, decelerates the bubble expansion, thus postponing the formation of the film of vapour. Therefore,  a hydrophilic surface provides better energy transfer from the hot wall to the water. 

Regardless of the surface topology and chemistry, we have shown that the maximum amount of energy transfer between the hot wall and the water increases with the wall temperature at the lowest temperature values considered (from 400~K to approximately 550~K depending on the wall topology). 
We explain this increase by quantifying the maximum rate of energy transfer, which is also increasing with the wall temperature.
Further increasing the wall temperature, the maximum amount of energy transfer undergoes a reduction. This reduction is a consequence of the formation of the vapour film which insulates the wall. Higher wall temperature accelerates bubble nucleation, bubble growth, and the formation of the film of vapour. Therefore, irrespective of the wall topology and chemistry,  we find a wall temperature for which the amount of transferred energy is maximum. 
Finally, our simulations show that although the averaged temperature of the system is always below the wall temperature, local temperature reaches that of the thermostat in the nucleation site and this hotter region grows in size as the bubble expands.

In summary, despite the known limitations of MD approach 
(e.g. short time scale, small systems, need for large superheat),
we show that large-scale MD simulations 
provide a viable tool to shed light on the combined effect 
of chemistry and nanostructure on the first stages of pool boiling.
The possibility to accurately control pressure, 
wall chemistry and nanostructure shape combined with the 
increasing computational performance of GPU systems  
pave the way to the use of this approach to explore 
more complex scenarios such as biphilic surfaces and reentrant
textures. 

\section*{Declaration of Competing Interest}
The authors declared that there is no conflict of interest.
\section*{Acknowledgement}
The research was financially supported by the Swedish Research Council, via the multidisciplinary research environment INTERFACE (VR  2016-06119 ``Hybrid multiscale modelling of transport phenomena for energy efficient processes"). The computation resources were supported by a grant from the Centro Svizzero di Calcolo Scientifico (CSCS) under project ID s864






\bibliographystyle{elsarticle-num} 
\bibliography{mybibfile}



\end{document}